\documentclass[12pt]{article}

\usepackage{sbc-template}
\usepackage{graphicx,url}
\usepackage[utf8]{inputenc}
\usepackage[english]{babel}

\usepackage{xspace}
\usepackage{amsfonts}
\usepackage{amssymb}
\usepackage{multirow}
\usepackage{amsfonts}
\usepackage{caption}
\usepackage{subfig}
\usepackage{listings, color}
\usepackage{xcolor}

\definecolor{bbeck}{RGB}{248, 118, 109}
\definecolor{dmuibum}{RGB}{163, 165, 0}
\definecolor{agracillis}{RGB}{0, 191, 125}
\definecolor{lamelatta}{RGB}{0, 176, 246}
\definecolor{octona}{RGB}{231, 107, 243}

\newcommand{\Red}[1]{{\color{dmuibum}#1}}
\newcommand{\Ora}[1]{{\color{bbeck}#1}}
\newcommand{\Blue}[1]{{\color{octona}#1}}
\newcommand{\Green}[1]{{\color{agracillis}#1}}
\newcommand{\Gray}[1]{{\color{lamelatta}#1}}

\def\helix{\texttt{HELIX}\xspace}

\title{\helix:\hspace{2px}A data-driven characterization of Brazilian land snails}

\author{Marcelo N. Almeida\inst{1}, Rodolfo Alves de Oliveira\inst{1}, Luiz Olmes\inst{2} \\ Gustavo S. Semaan\inst{1}, Daniel de Oliveira\inst{3}, Lúcio Santos\inst{4}, and Marcos Bedo\inst{1}}

\address{Fluminense Northwest Institute -- Fluminense Federal University (UFF), Brazil 
\email{\{mnocelle, rodolfooliveira, gustavosemaan, marcosbedo\}@id.uff.br}
\nextinstitute Inst. of Comp. and Mathematics --  Federal University of Itajubá (UNIFEI), Brazil \email{olmes@unifei.edu.br} 
\nextinstitute Institute of Computing -- Fluminense Federal University (UFF), Brazil 
\email{danielcmo@ic.uff.br}
\nextinstitute Federal Institute of North of Minas Gerais (IFNMG), Brazil
\email{lucio.santos@ifnmg.edu.br}
}

\begin{document} 

\noindent
\textbf{DISCLAIMER:} 
This paper has been accepted for publication in the Proceedings of the XXXVI Brazilian Symposium on Databases -- (SBBD'21) -- Short Papers -- Online. SBC. \url{https://sbbd.org.br/2021/}

\newpage

\maketitle

\begin{abstract}
Decision-support systems benefit from hidden patterns extracted from digital information.
In the specific domain of gastropod characterization, morphometrical measurements support biologists in the identification of land snail specimens.
Although snails can be easily identified by their excretory and reproductive systems, the after-death mollusk body is commonly inaccessible because of either soft material deterioration or fossilization. 
This study aims at characterizing Brazilian land snails by morphometrical data features manually taken from the shells.
In particular, we examined a dataset of shells by using different learning models that labeled snail specimens with a precision up to $97.5\%$ (F1-Score = $.975$, CKC = $.967$ and ROC Area = $.998$).
The extracted patterns describe similarities and trends among land snail species and indicates possible outliers physiologies due to climate traits and breeding. 
Finally, we show some morphometrical characteristics dominate others according to different feature selection biases. 
Those data-based patterns can be applied to fast land snail identification whenever their bodies are unavailable, as in the recurrent cases of lost shells in nature or private and museum collections.
\end{abstract}

\section{Introduction}

While land snails are a recurrent presence within the Brazilian tropical fauna scene, their identification and labeling are still a challenge since snail bodies quickly deteriorate after death.
On the other hand, the shells, which are made of calcium carbonate, remain preserved long after.
Accordingly, land snail shells are easily found in the wild and museums, and their investigation is one of the main sources in the characterization of such a class of gastropods~\cite{Ueta1980,Slapcinsky2016}.
For instance, chemical analyses enable the extraction of climate, food, and genomics information, usually destroying (parts of) the shell in the process~\cite{Simone2006,Richling2021}. 

A classical, non-invasive approach to label those shells is to
\textit{(i)}~portray their physiology and then
\textit{(ii)}~compare the collected measures against a \textit{golden} pattern or reference~\cite{Simone2006,De2007,Slapcinsky2016}.
This \underline{morphometrical comparison process} enables the fast identification of shells within collections and provides a rationale to find possible related species by similarity~\cite{Quenu2020,Yeung2020}.
In this study, we take a \underline{step further} to enhance such a comparison-based process by \underline{extracting data-driven patterns} from different learning models tested over a handcrafted dataset of Brazilian land snail shells.

\begin{figure}[!t]
\centering
\subfloat{%
  \includegraphics[width=.313\textwidth]{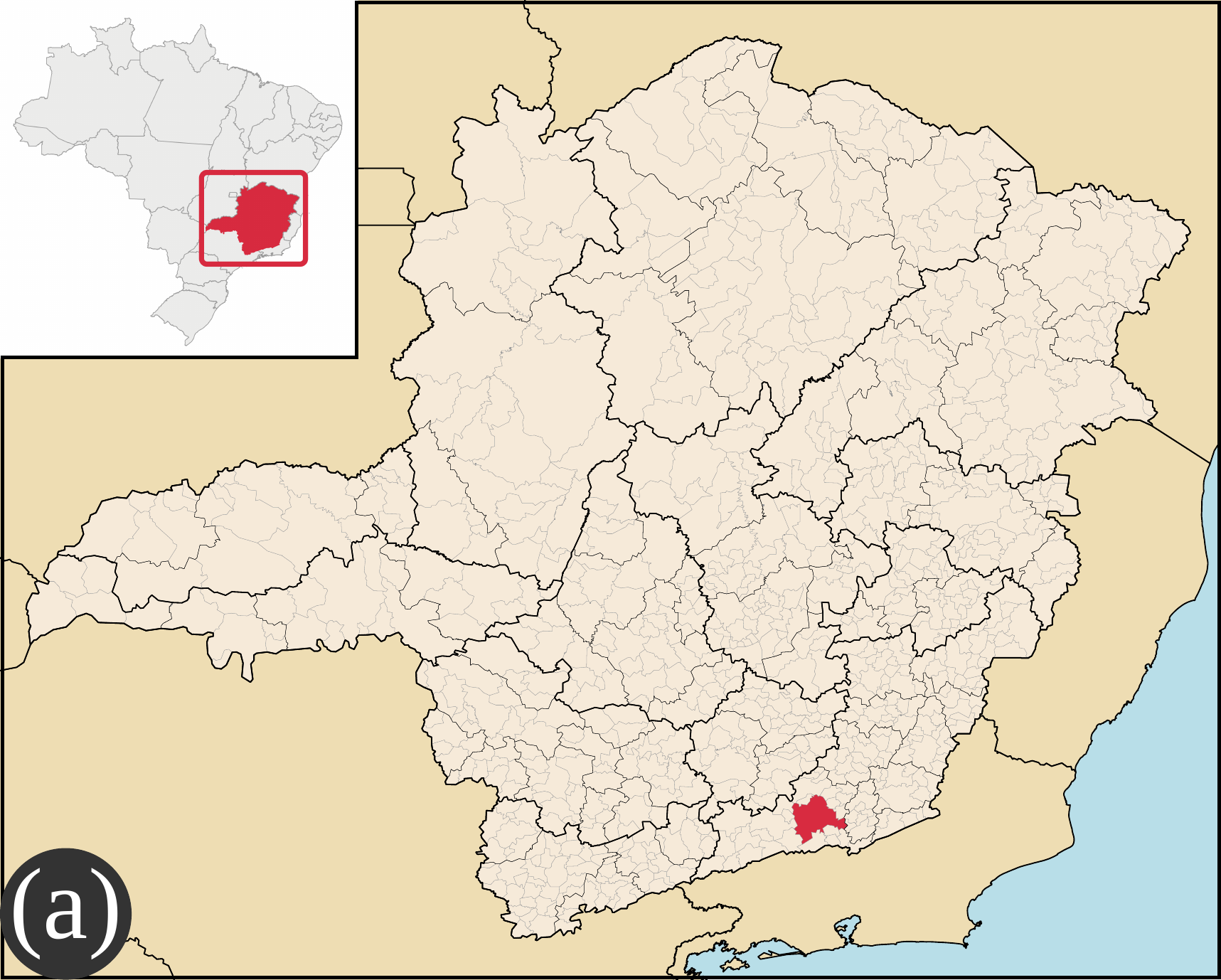}\label{fig:juiz_de_fora}
}~
\subfloat{%
  \includegraphics[width=.61\textwidth]{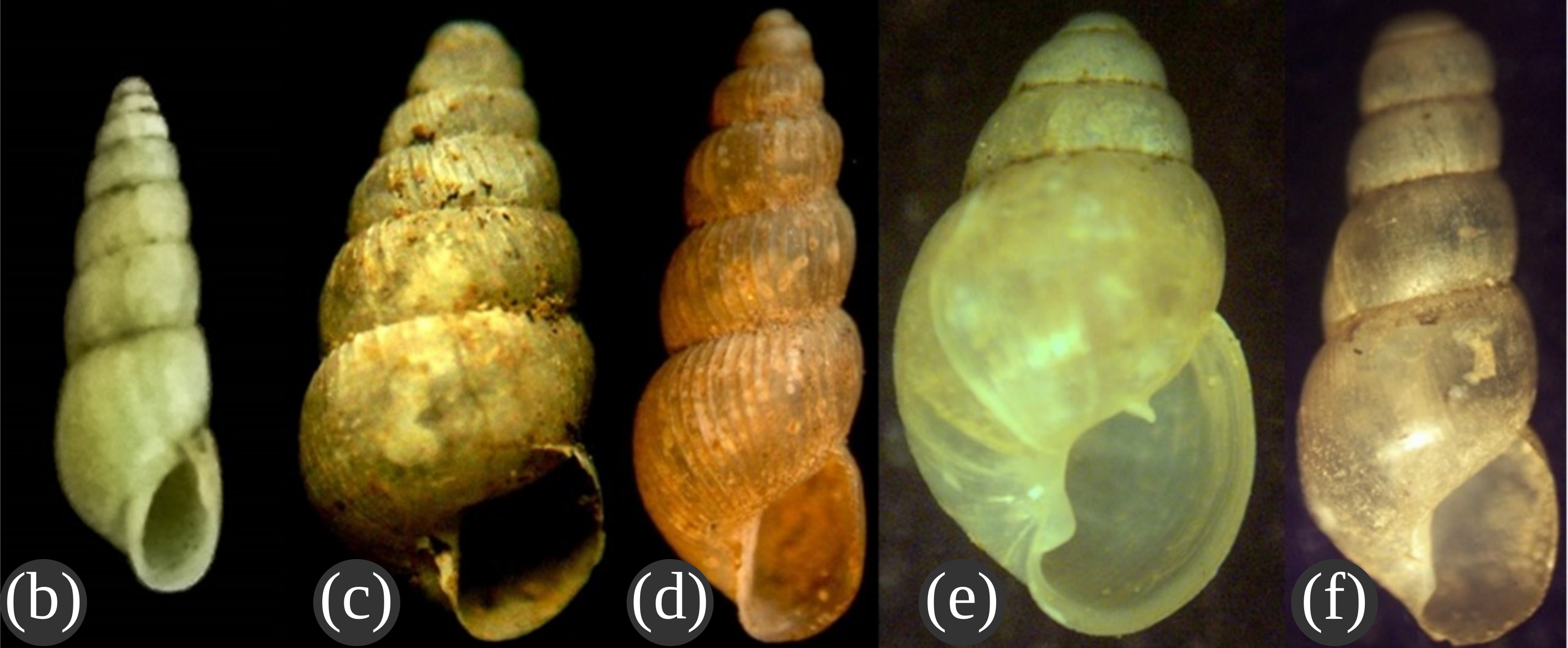}\label{fig:classes_conchas}%
}
\caption{Brazilian land snails specimens.
(a)~Shells found at the Juiz de Fora/MG metro area.
(b)~\Green{\textit{Allopeas gracilis}}. 
(c)~\Ora{\textit{Beckianum beckianum}}.
(d)~\Red{\textit{Dysopeas muibum}}.
(e)~\Gray{\textit{Leptinaria unilamellata}}.
(f)~\Blue{\textit{Subulina octona}.}}
\end{figure}

In particular, we curated a dataset (coined \helix)\footnote{{{\footnotesize{\helix dataset is available at \url{zenodo.org/record/5500215#.YUCgD51Kg2w}.}}}} that contains shells gathered at the metro area of the Brazilian city of Juiz de Fora/MG, whose land snail species are vastly documented\footnote{See {{\footnotesize\url{ufjf.br/malacologia}}}}.
Figure~\ref{fig:juiz_de_fora} shows the geographical location of shells, while Figure~\ref{fig:classes_conchas} presents a representative shell instance for each of the five different \helix species.
Every shell was manually collected throughout a year, and the dataset covers a wide variety of individual mollusks of different genders and ages.
Accordingly, we took full advantage of the KDD pipeline for mining \helix data~\cite{Aggarwal2015}.

Pre-processed instances were feed to fine-tuned Na\"ive-Bayes, JRip, Decision Tree, RandomForest, Instance-based Learning, Functional-Tree, and Multi-Layer Perceptron (MLP) classifiers, which excelled in the labeling task with a precision up to $97.5\%$.
Accordingly, we examined the learned models and found patterns showing 
\textit{(i)}~a geometric separation between the specimens,
\textit{(ii)}~similar land snails according to their shell morphometrics,
\textit{(iii)}~correlations between hierarchical characteristics and the snail species,
\textit{(iv)}~rules for associating a subset of shell features to the species, and
\textit{(v)}~the dominance of certain characteristics in the labeling processes.
Such data-driven outcomes provide insights to the comparison process by explicitly describing the morphometrical patterns of land snail shells as well as highlighting spatial similarity trends.

The remainder of this paper is organized as follows.  
Section~2 provides the background and related work, while Section~3 describes the material and methods.
Finally, Sections~4 and 5 provide the experimental evaluation and conclude the study.

\section{Preliminaries and Related Work}

The morphometrical comparison process of land snails relies on portraying and measuring the snail shells for further matching against baselines of independent and correlated morphological characteristics.
For instance, the study of Ueta (1980) provides a morphometrical catalog for Brazilian land snails.
The comparison process can also be coupled with genomics to 
\textit{(i)}~expand current taxonomies and 
\textit{(ii)}~identify relationships between extant and extinct species~\cite{De2007}.
For instance, the study of Hirano \textit{et. al}~(2018) uses morphometric traits combined with phylogenetic relationships to demonstrate an ocean snail considered extinct was still extant.
Yeung \textit{et. al}~(2020) also relied on morphometrics to introduce a new Hawaiian snail that was not related to any known local species.
Analogously, the study of Slapcinsky and Kraus (2016) uses a morphometrical comparison to review the existing Palaus species and identify new morphological characteristics.

Recently, learning models have been used to enhance and speed up the comparison process, as in the study of Quenu \textit{et. al}~(2020).
The authors employed two learning models for the identification of phenotypes from New Caledonia snail species: an MLP classifier and a Gaussian mixture model.
The results indicated a proportion of the $1/3$ of individuals was in an overlapped cluster among the patterns of two species.

\section{Material and Methods}

\begin{figure}[!t]
\begin{tabular}{ll}
\begin{tabular}{c}
\includegraphics[width=.17\textwidth]{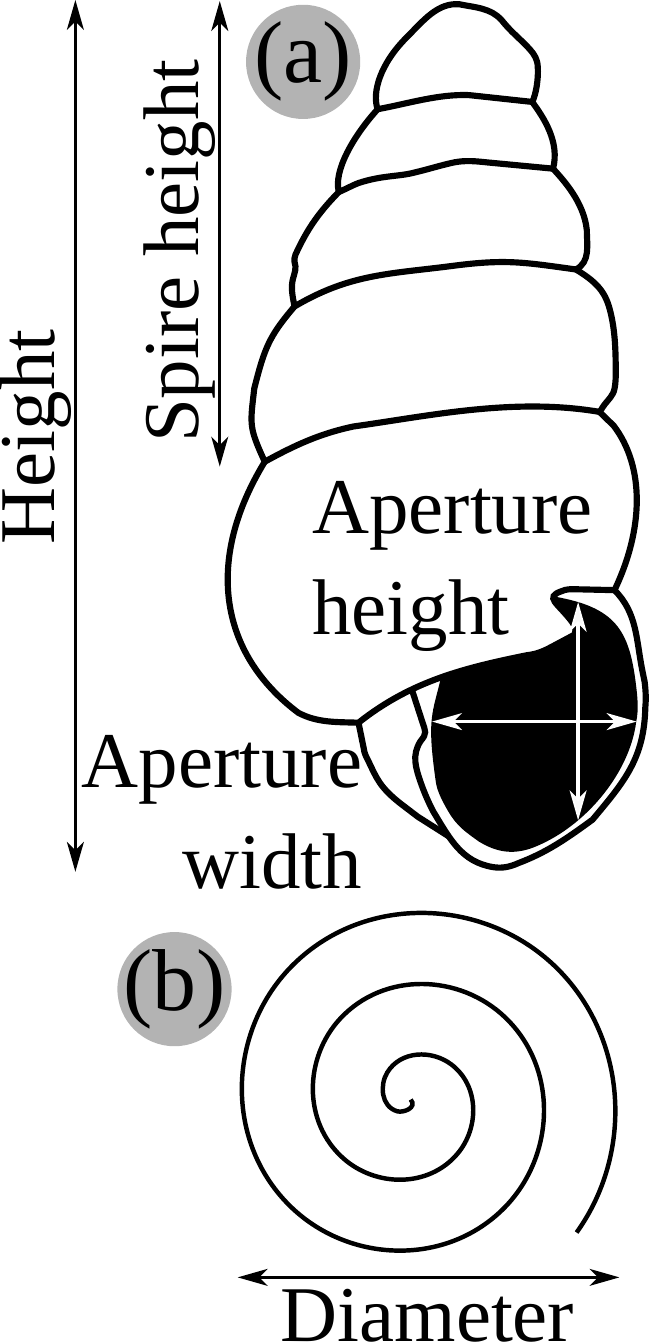}
\end{tabular}
& 
{{\footnotesize
\begin{tabular}{l|l}\hline\hline
\textbf{Rule $\downarrow$ (All measures are in mm.)} & \textbf{Species} \\ \hline
Aperture height {{\tiny$\leq$}} 2.1 {{\tiny$\land$}} Diameter {{\tiny$\leq$}} 2.1  &  \Red{D. muibum} \\ \hline
Aperture height {{\tiny$>$}} 2.1 {{\tiny$\land$}} Diameter {{\tiny$>$}} 4.7 & \Gray{L. unilamel.}   \\ \hline
Aperture height {{\tiny$\leq$}} 2.1 {{\tiny$\land$}} 2.1 {{\tiny$<$}} Diameter {{\tiny$\leq$}} 2.4 {{\tiny$\land$}} Height {{\tiny$\leq$}} 4.8  & \Ora{B. beckianum} \\ \hline
Aperture height {{\tiny$\leq$}} 2.1 {{\tiny$\land$}} 2.1 {{\tiny$<$}} Diameter {{\tiny$\leq$}} 2.4 {{\tiny$\land$}} Height {{\tiny$>$}} 4.8 & \Blue{S. octona} \\ \hline
Aperture height {{\tiny$\leq$}} 2.1 {{\tiny$\land$}} Diameter {{\tiny$>$}} 2.4 & \Ora{B. beckianum} \\ \hline
2.1 {{\tiny$<$}} Aperture height {{\tiny$\leq$}} 2.9  {{\tiny$\land$}} Diameter {{\tiny$\leq$}} 4.7   & \Blue{S. octona} \\ \hline
Aperture height {{\tiny$>$}} 2.9 {{\tiny$\land$}} Diameter {{\tiny$\leq$}} 4.7 {{\tiny$\land$}} Spire height {{\tiny$\leq$}} 5.8  & \Green{A. gracilis}  \\ \hline
Ap. ht. {{\tiny$\leq$}} 3.1 {{\tiny$\land$}} Diameter {{\tiny$\leq$}} 3.6 {{\tiny$\land$}} Spire height {{\tiny$>$}} 5.8 & \Blue{S. octona} \\ \hline
3.1 {{\tiny$<$}} Ap. ht. {{\tiny$\leq$}} 3.5 {{\tiny$\land$}} Dia. {{\tiny$\leq$}} 3.6 {{\tiny$\land$}} S. ht. {{\tiny$>$}} 5.8 {{\tiny$\land$}}  Ht. {{\tiny$\leq$}} 11.2  &  \Green{A. gracilis} \\ \hline
3.1 {{\tiny$<$}} Ap. ht. {{\tiny$\leq$}} 3.3 {{\tiny$\land$}} Dia. {{\tiny$\leq$}} 3.6 {{\tiny$\land$}} S. ht. {{\tiny$>$}} 5.8 {{\tiny$\land$}}  Ht. {{\tiny$>$}} 11.2  & \Blue{S. octona} \\ \hline
3.3 {{\tiny$<$}} Ap. ht. {{\tiny$\leq$}} 3.5 {{\tiny$\land$}} Dia. {{\tiny$\leq$}} 3.6 {{\tiny$\land$}} S. ht. {{\tiny$>$}} 5.8 {{\tiny$\land$}}  Ht. {{\tiny$>$}} 11.2   & \Green{A. gracilis} \\ \hline
2.9 {{\tiny$<$}} Ap. ht. {{\tiny$\leq$}} 3.5 {{\tiny$\land$}} 3.6 {{\tiny$<$}} Diameter {{\tiny$\leq$}} 4.7 {{\tiny$\land$}} S. ht. {{\tiny$>$}} 5.8  & \Blue{S. octona} \\ \hline
Ap. ht. {{\tiny$>$}} 3.5 {{\tiny$\land$}} Diameter {{\tiny$\leq$}} 4.7 {{\tiny$\land$}} 5.8 {{\tiny$<$}} Spire height {{\tiny$\leq$}} 9.9 & \Green{A. gracilis}  \\ \hline
Ap. ht. {{\tiny$>$}} 3.5 {{\tiny$\land$}} Diameter {{\tiny$\leq$}} 4.7 {{\tiny$\land$}} Spire height {{\tiny$>$}} 9.9 & \Blue{S. octona}  \\ \hline
\hline
\end{tabular}
}}
\end{tabular}
\caption{Schema for (a)~vertical and (b)~horizontal morphometrical shell features, and generative rules (induced by Figure~\ref{fig:DT}) for the \helix dataset.}\label{fig:esquema} 
\end{figure}

\noindent
\textbf{\helix dataset.}
The \textit{Subulinidae} family of Brazilian land snails includes similar species with particular morphometrical characteristics.
We surveyed the metro area of the Juiz de Fora/MG city and collected snails from a representative private property with $62{,}000$m$^2$ centered at $21^{o}42'31''S$, $43^{o}21'26''W$, alt. 795m, with red-yellow oxisol, 8.5 pH, and Cwa climate.
We performed monthly collections in a ecosystem with \textit{Pennisetum purpureum}, \textit{Brachiaria mutica}, \textit{Paspalum notatum}, \textit{Bidens pilosa}, \textit{Leucena leucocephala}, and \textit{Ricinus communis} plants near a river flow in the 2008/Sep -- 2009/Aug timespan.
There, we defined an equally spaced transect of 200m with ten collection points and gathered $50\times50$cm, $500$g litter-falls.
Samples were sieved by $2.0$mm meshs, and living specimens were cleaned and fixed in a Railliet-Henry liquid, their shells removed, dried, and separated. 
Finally, we followed the guidelines of Simone (2006) to label the species and create the \helix dataset.
We measured each shell \underline{Height}, \underline{Diameter}, \underline{Spire height}, and \underline{Aperture width} and \underline{height} with a pachymeter.
Figures~\ref{fig:esquema}a--b show the representation of the characteristics.
The dataset includes $518$ instances labeled as \Ora{B. beckianum} ($28.8\%$), \Red{D. muibum} ($21.2\%$), \Green{A. gracilis} ($9.7\%$), \Gray{L. unilamellata} ($12\%$), and \Blue{S. octona} ($28.3\%$).

\noindent
\textbf{Tools and learning models.}
We examined \helix data by using seven classifiers of distinct paradigms implemented by the Weka workbench\footnote{Available at: {{\small\url{cs.waikato.ac.nz/ml/weka/}}}} coded in Java and binded into R $3.6.3$ v(2020-02-29) through package `RWeka'\footnote{Available at: {{\small\url{cran.r-project.org/web/packages/RWeka/index.html}}}}.
We also implemented the intrinsic dimensionality estimator~\cite{Chavez2001} for PCA transformation and feature selection in R.
Feature selection procedures were implemented according to the Pearson, Relieff, and Fast Correlation-Based Filter (FCFB) criteria~\cite{Roffo2017} with individual rankings aggregated with the MedianRank algorithm~\cite{Fagin2003}. 
As a result, we tuned the classifiers by using the original \helix data and its versions with 
\textit{(i)}~selected features and 
\textit{(ii)}~PCA-transformed instances.
Each of the three data versions was scaled and feed in different and isolated \textit{batches} to the models following a $10$-fold cross-validation routine.
We experimented with a broad set of parameters aiming to avoid overfitting and repeated the execution of randomized approaches $10\times$ with distinct seeds. 

\section{Experiments}

\begin{table}[!t]
\caption{Feature selection with MedianRank aggregation.}
\footnotesize
\centering 
\begin{tabular}{c|c|c} 
\hline\hline 
\textbf{Pearson $\downarrow$} & \textbf{Relieff $\downarrow$} & \textbf{FCBF $\downarrow$} \\  \hline
\hline
\textit{Aperture ht.} $(.47)$ & \textit{Aperture ht.} $(.23)$  & Diameter $(.78)$     \\
\textit{Height} $(.42)$       & \textit{Height} $(.17)$        & \textit{Height} $(.58)$       \\
Spire ht. $(.39)$    & Aperture wt. $(.16)$  & \textit{Aperture ht.} $(.53)$ \\
Aperture wt. $(.37)$ & Diameter $(.14)$      & Spire ht. $(.46)$    \\
Diameter $(.27)$ & Spire ht. $(.11)$         & Aperture wt. $(.45)$ \\ \hline 
\hline 
\end{tabular}
\label{tab:feature_selection}
\end{table}
\begin{table}[!t]
\caption{Consolidated best/worst results for all batches and setups.}
\footnotesize
\centering 
\begin{tabular}{c|c|c|c|l} 
\hline\hline 
 \textbf{Classifier $\downarrow$} & \textbf{F1} & \textbf{CKC} & \textbf{ROC} & \textbf{Parameters} \\  \hline
\hline
Na\"ive-Bayes      & .870/.792  & .827/.768 & .981/.941 & Normal, non-conditional d.p.f. \\ \hline 
JRip               & .936/.859  & .917/.818 & .982/.947 & 03 folds rule-finding  \\ \hline 
Decision-Tree      & .947/.835  & .932/.791 & .975/.945 & Pruned, n-ary tree w/ entropy   \\ \hline 
RandomForest       & .952/.864  & .939/.823 & .996/.967 & 100 iterations, 30-inst. per bag  \\ \hline 
IB Learning        & .965/.850  & .955/.806 & .986/.945 & 03 neighbors w/ $L_2$ distance   \\ \hline 
Functional-Tree    & .968/.855  & .959/.817 & .994/.961 & Oblique tree w/ linear function   \\ \hline\hline
ML Perceptron      & .975/.858  & .967/.822 & .998/.954 & 01 hidden layer w/ 05 neurons   \\ \hline 
\hline 
\end{tabular}
\label{tab:classifiers}
\end{table}


We constructed three test batches representing the
\textit{(i)}~original instances,
\textit{(ii)}~instances with selected features, and 
\textit{(iii)}~PCA-transformed data by using the rounded \helix \underline{intrinsic} \underline{dimension (equals 2)}.
Feature selection was carried out with three filters and aggregated with MedianRank algorithm, as described in Table~\ref{tab:feature_selection}.
The characteristics \textit{Aperture height} and \textit{Height} were the two most dominant and, consequently, were used to construct the second batch of \helix tests.
Table~\ref{tab:classifiers} shows the best and worst performances achieved by the classifiers and their tuning parameters for each batch test.
Results indicate the labeling-driven learning models are suitable to the shell identification problem, as their performances were bounded into the $[.792, .975]$ F1--Score interval.
Accordingly, we examined every learned model to describe the patterns and their biological interpretations.

\begin{figure}[!t]
\centering
\includegraphics[width=0.85\textwidth]{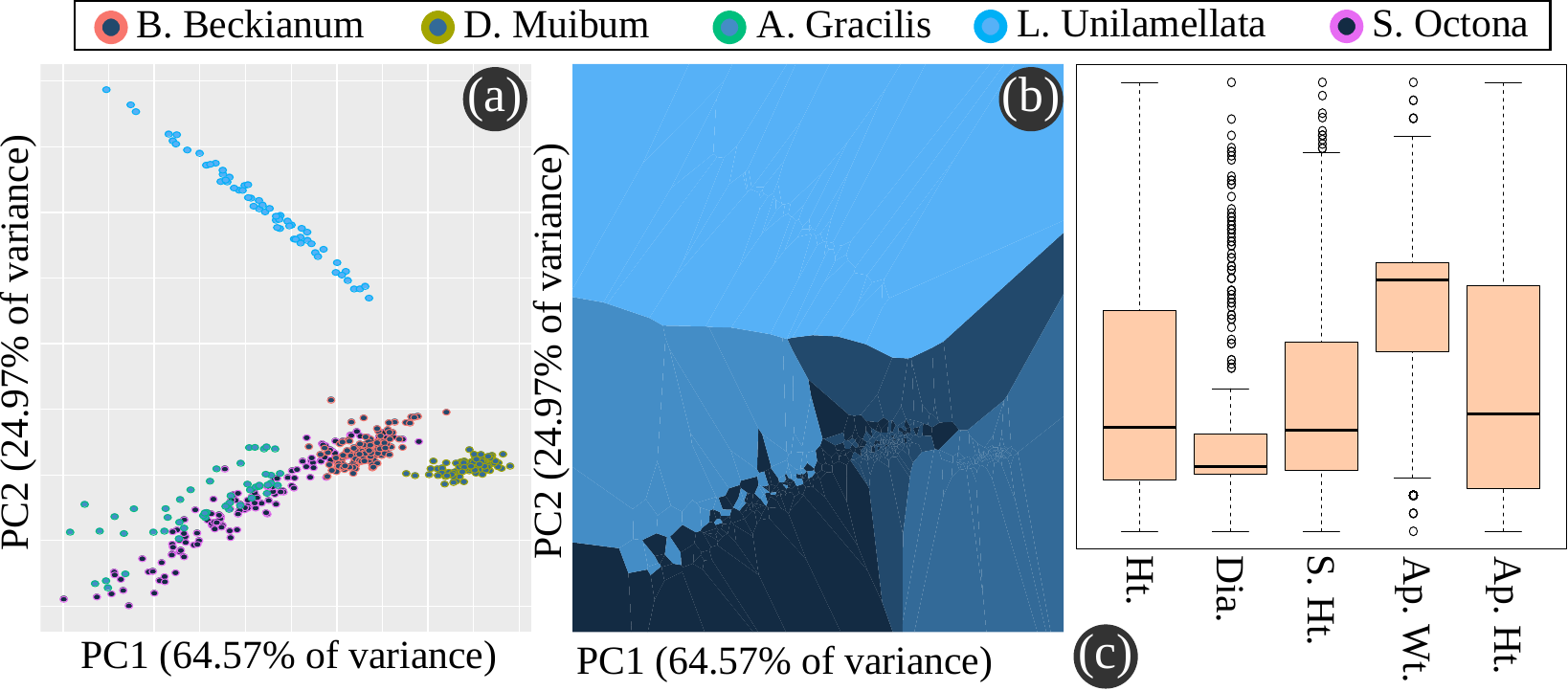}\label{fig:voronoi}
\caption{(a)~PCA data visualization, (b)~Voronoi-based similarity coverage for each land snail, and (c)~$[0,1]$-scaled box-plot of \helix attributes.}\label{fig:similarity}
\end{figure}
\begin{figure}[!t]
\centering
\includegraphics[width=\textwidth]{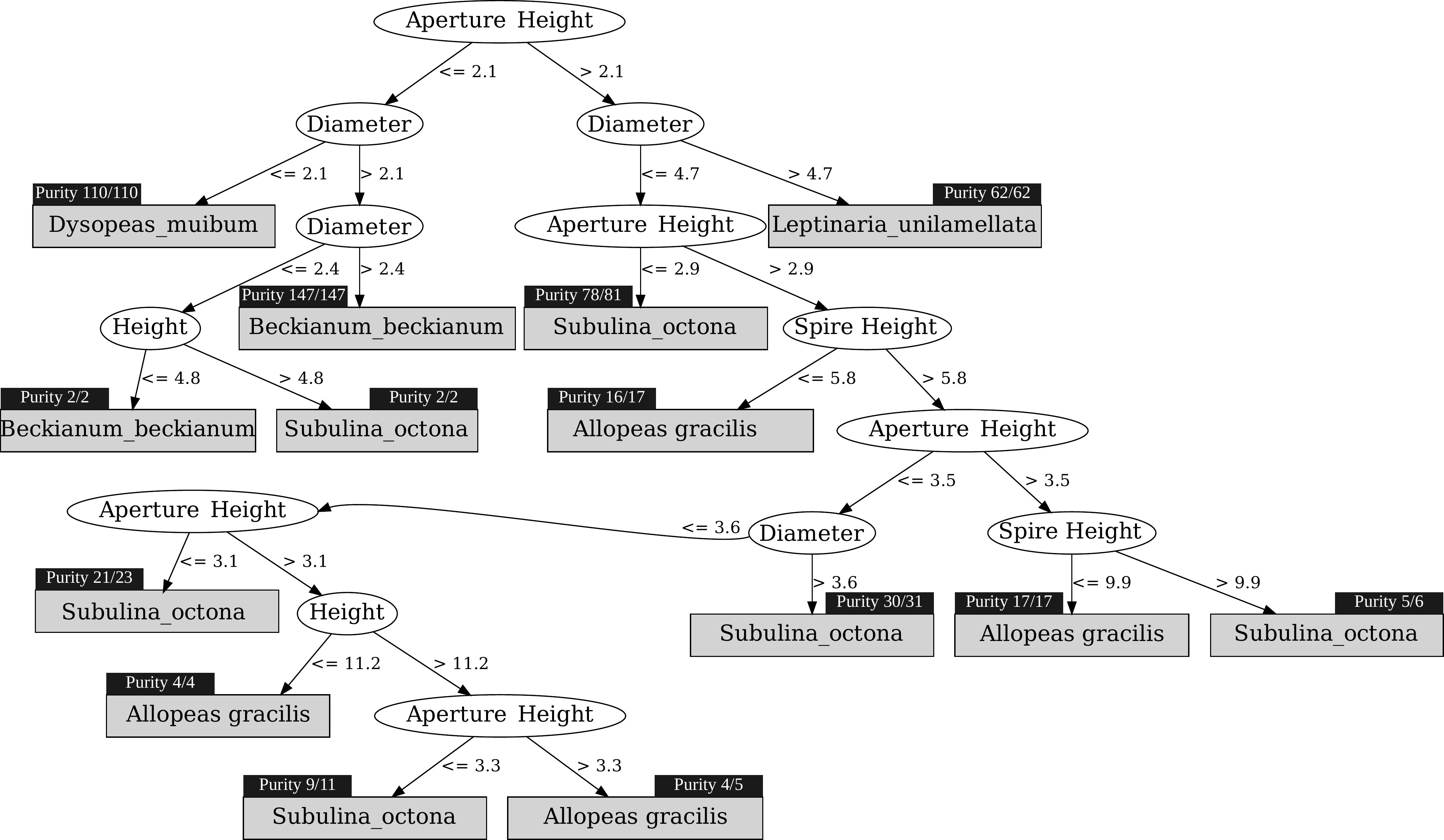}
\caption{Decision-Tree constructed by the C4.5 learning model.}\label{fig:DT}
\end{figure}

\noindent
\textbf{Distance-based relationships and boundaries.}
The PCA visualization in Figures~\ref{fig:similarity}a--b show \Gray{L. unilamellata} and \Red{D. muibum} are spatially separated and clustered in Euclidean-based Voronoi spaces with soft borders, which is biologically explained by morphometrics, \textit{i.e.}, the former land snails have \underline{bulkier shells} and the latter \underline{shortened ones}.
The results also indicate two cluster overlappings among \Ora{B. beckianum}, \Blue{S. octona} and \Green{A. gracilis} specimens, which is explained by the \underline{higher phylogenetical proximity} among those species.
Neither of those observations can be made only by checking individual nor pairwise attributes -- Figure~\ref{fig:similarity}c.


\noindent
\textbf{Decision-Tree and association rules.}
The most frequent decision-tree constructed after the cross-validation procedure in Table~\ref{tab:classifiers} is presented in Figure~\ref{fig:DT}. 
It describes 
\textit{(i)}~a hierarchical correlation of attributes with shorter paths for describing \Gray{L. unilamellata} and \Red{D. muibum} (corroborating Figure~\ref{fig:similarity}b), and
\textit{(ii)}~a generative rule set that is detailed in Figure~\ref{fig:esquema}.
The decision root node is the \textit{Aperture height} (matching Table~\ref{tab:feature_selection}), and the tree paths show the most internal species is the \Green{A. gracilis}, which can only be reached if at least three attributes are determined.
Results also show the densest and pure nodes are in the first two levels, indicating \textit{Height} and \textit{Diameter} are used to separated species' inner borders.
Such findings are biologically interpreted as \Ora{B. beckianum}, \Blue{S. octona} and \Green{A. gracilis} specimens having similar \textit{Aperture height} and \textit{Diameter} \underline{depending} on their \underline{age} and \underline{eco-climate traits}.
Finally, generative rules constructed after the Decision-Tree have shown support of at least $80\%$, which indicates those data-based patterns can be applied for fast land snail identification whenever their bodies are unavailable, as in the recurrent cases of lost shells in nature, private, or museum collections.

\section{Conclusions and Future Work}

This study has discussed a data-driven characterization of Brazilian land snails.
In particular, we have 
\textit{(i)}~curated a reference dataset by manually gathering and measuring a set of land snail shells at the metro area of the Brazilian city of Juiz de Fora/MG and
\textit{(ii)}~evaluated those data by different learning models.
Results indicate classifier models labeled land snail shells with a precision up to $97.5\%$ (F1--Score $= .975$, CKC $= .967$, and ROC Area $= .998)$, whereas learned models presented patterns regarding the proximity of species and the generative rules for the specimens.
Such outcomes provide insights into the land morphometrical comparison process by describing data-driven trends.
Future works include the \helix extension to include other Brazilian land snail species.

\noindent
\textbf{Acknowledgments.} The study was supported by CNPq, CAPES, and FAPERJ.

\bibliographystyle{sbc}
\bibliography{sbc-template}

\begin{thebibliography}{}

\bibitem[Aggarwal 2015]{Aggarwal2015}
Aggarwal, C. (2015).
\newblock {\em {Data mining: The textbook}}.
\newblock Springer.

\bibitem[Ch{\'a}vez et~al. 2001]{Chavez2001}
Ch{\'a}vez, E., Navarro, G., Baeza-Yates, R., and Marroqu{\'\i}n, J. (2001).
\newblock Searching in metric spaces.
\newblock In {\em Computing Surveys}, volume~33, pages 273--321. ACM.

\bibitem[Fagin et~al. 2003]{Fagin2003}
Fagin, R., Kumar, R., and Sivakumar, D. (2003).
\newblock Efficient similarity search and classification via rank aggregation.
\newblock In {\em ACM SIGMOD}, pages 301--312.

\bibitem[Hirano et~al. 2018]{Richling2021}
Hirano, T., Wada, S., Mori, H., Uchida, S., Saito, T., and Chiba, S. (2018).
\newblock Genetic and morphometric rediscovery of an extinct land snail on
  oceanic islands.
\newblock {\em J. of Molluscan S.}, 84(2):148--156.

\bibitem[Queiroz 2007]{De2007}
Queiroz, K. (2007).
\newblock Species concepts and species delimitation.
\newblock {\em Sys. Bio.}, 56(6):879--886.

\bibitem[Quenu and \textit{et. al} 2020]{Quenu2020}
Quenu, M. and \textit{et. al} (2020).
\newblock {Geometric morphometrics and ML challenge currently accepted species
  limits of the land snail \textit{Placostylus}}.
\newblock {\em J. M. Studies}, 86(1):35--41.

\bibitem[Roffo 2017]{Roffo2017}
Roffo, G. (2017).
\newblock Ranking to learn and learning to rank: On the role of ranking in
  pattern recognition applications.
\newblock {\em arXiv preprint arXiv:1706.05933}.

\bibitem[Simone 2006]{Simone2006}
Simone, L. R. L.~d. (2006).
\newblock {\em Land and freshwater molluscs of Brazil}.
\newblock Museu de Zoologia, Universidade de S{\~a}o Paulo.

\bibitem[Slapcinsky and Kraus 2016]{Slapcinsky2016}
Slapcinsky, J. and Kraus, F. (2016).
\newblock Revision of partulidae of {Palau}, with description of a new genus
  for an unusual ground-dwelling species.
\newblock {\em ZooKeys}, 86(614):27.

\bibitem[Ueta 1980]{Ueta1980}
Ueta, M.~T. (1980).
\newblock Estudo morfom{\'e}trico da concha de \textit{Lymnaea columella}.
\newblock {\em R. Soc. Bras. Medicina Tropical}, 13(1):119--141.

\bibitem[Yeung and \textit{et. al} 2020]{Yeung2020}
Yeung, N. and \textit{et. al} (2020).
\newblock Overlooked but not forgotten: the first new extant species of
  hawaiian land snail described in 60 years, \textit{Auriculella gagneorum}.
\newblock {\em ZooKeys}, 950:1.

\end{thebibliography}

\end{document}